\documentclass[a4paper]{article}

\usepackage{INTERSPEECH2018}

\title{Extending Recurrent Neural Aligner for Streaming End-to-End Speech Recognition in Mandarin}
\name{Linhao Dong$^{1,2}$, Shiyu Zhou$^{1,2}$, Wei Chen$^1$, Bo Xu$^1$
}
\address{
$^1$Institute of Automation, Chinese Academy of Sciences, China\\
$^2$University of Chinese Academy of Sciences, China}
\email{\{donglinhao2015, zhoushiyu2013, w.chen, xubo\}@ia.ac.cn}

\begin{document}
\maketitle
\begin{abstract}
End-to-end models have been showing superiority in Automatic Speech Recognition (ASR). At the same time, the capacity of streaming recognition has become a growing requirement for end-to-end models. Following these trends, an encoder-decoder recurrent neural network called Recurrent Neural Aligner (RNA) has been freshly proposed and shown its competitiveness on two English ASR tasks. However, it is not clear if RNA can be further improved and applied to other spoken language. In this work, we explore the applicability of RNA in Mandarin Chinese and present four effective extensions: In the encoder, we redesign the temporal down-sampling and introduce a powerful convolutional structure. In the decoder, we utilize a regularizer to smooth the output distribution and conduct joint training with a language model. On two Mandarin Chinese conversational telephone speech recognition (MTS) datasets, our Extended-RNA obtains promising performance. Particularly, it achieves 27.7\% character error rate (CER), which is superior to current state-of-the-art result on the popular HKUST task.


\end{abstract}

\section{Introduction}
Recently, a considerable amount of works have demonstrated the simplification and effectiveness of end-to-end models in the ASR field \cite{graves2012sequence, chan2016listen, jaitly2016online, raffel2017online, sak2017recurrent}. These models eschew the needs of finite state transducers (FST), pronouncing lexicons or any expert knowledge in the conventional ASR systems and directly recognize speech utterance by a single neural network.

Among these models, Listen, Attend and Spell (LAS) \cite{chan2016listen} is a very popular attention-based model, which achieves delightful improvements over the state-of-the-art conventional ASR systems \cite{chiu2017state}. However, the limitation that the input sequence must be entirely encoded before conducting attention hampers LAS in supporting streaming recognition. Neural Transducer (NT) \cite{jaitly2016online} and Monotonic Alignments \cite{raffel2017online} are therefore proposed to revise the LAS model with online attention mechanisms, and form one branch of streaming end-to-end models.

Besides above attention-based models, Recurrent Neural Network Transducer (RNN-T) \cite{graves2012sequence} and Recurrent Neural Aligner (RNA) \cite{sak2017recurrent} are also end-to-end models within the encoder-decoder framework. They extend the CTC \cite{graves2006connectionist} approach and produce strictly monotonic alignments between the input sequence and target sequence. Unlike the CTC model, they don't make a conditional independence assumption between predictions at different steps. Specifically, RNN-T keeps alternating between updating the transcription and the prediction network based on if the predicted label is a blank or not. RNA is simpler than RNN-T and encodes the predicted label as a input to the decoder no matter what the label is. These two models form another branch of end-to-end models offer streaming decoding.

%


Comparing with other streaming end-to-end models, RNA has following advantages: (1) RNA produces left-to-right alignments naturally, which has less building complexity than online attention-based models; (2) RNA predicts one output label at each time step in input, rather than multiple labels by RNN-T, thus simplifying the beam search decoding and making training more efficient. However, RNA has only been evaluated its effectiveness on English ASR task, and it is not clear if RNA can be successfully applied to other spoken language.

In this work, we introduce the RNA model to Mandarin Chinese, which has some noteworthy distinctions from English: (1) Mandarin has a lower temporal entropy density and less number of characters per second \cite{amodei2016deep}, thus may suitable for different temporal down-sampling mechanisms from English; (2) Each character of Mandarin has it own tonal pronunciation, which needs to be carefully distinguished during recognizing; (3) Mandarin has a larger alphabet set (commonly several thousands) than English, and most of characters have many homonyms, which are easy to be wrongly written. Based on above analyses, we extend RNA in following aspects:

\begin{itemize}
\vspace{-0.05cm}
\item We explore various temporal down-sampling mechanisms for Mandarin. The results show that 1/8 down-sampling rate performs best, and suitable combination of different down-sampling methods also matters.
\vspace{-0.05cm}
\item In order to capture more distinguishable acoustic details, we introduce Multiplicative Unit (MU) \cite{kalchbrenner2016video} to the encoder of RNA. And find it offers consistent CER reduction in Mandarin recognition.
\vspace{-0.05cm}
\item Since Mandarin has a large character set where most characters have various homonyms, we encourage more smooth output distribution by conducting Confidence Penalty\cite{pereyra2017regularizing}, which helps RNA to explore more sensible alternatives and obtain better generalization.
\vspace{-0.05cm}
\item For alleviating the wrongly written phenomenon of homonyms in Mandarin, we propose a joint training mechanism of RNA and Recurrent Neural Network Language Model (RNN-LM). The mechanism is designed for sidestepping the disturbance of blank label in RNA and could facilitate better recognition results.
\vspace{-0.05cm}
\end{itemize}
Our Extended-RNA model obtains competitive performance on both of two Mandarin Chinese ASR datasets. Especially, it establishes a new state-of-the-art CER performance of 27.7\% on the Mandarin Chinese ASR benchmark (HKUST) \cite{liu2006hkust}.


\section{Recurrent Neural Aligner}
Recurrent Neural Aligner (RNA) utilizes an encoder-decoder framework. Let $\bold{x}$ = ($x_1, x_2, ..., x_T$) be the input sequence of audio frame features and $\bold{y}$ = ($y_1, y_2, ..., y_N$) be the target sequence of labels. The encoder, which could be any neural network, transforms the original input sequence $\bold{x}$ into a high level representation $\bold{h}$ = ($h_1, h_2, ..., h_U$) with length $U \le T$:
\begin{equation}
  \bold{h} = \text{encoder}(\bold{x})
  \label{eq1}
\end{equation}
The decoder is a recurrent neural network with a softmax output layer containing $L+1$ units, where $L$ is the number of real labels and the additional one is the blank label. We simplify the original RNA decoder in \cite{sak2017recurrent} for lighter training and computation consistency between training and inference. Specifically, at step $u$, the input to decoder is the concatenation of encoder output $h_{u}$ and encoded vector $e_{u-1}$ of label $z_{u-1}$, which is the predicted label with the maximum probability among the softmax outputs at step $u-1$. So, the calculation of label $z_{u}$ can be formulated as:
\begin{equation}
  z_u = \mathop{\arg\max}_{l \in [1, L+1]}(\text{decoder}(h_u, e_{u-1}))
  \label{eq2}
\end{equation}
RNA defines a conditional distribution $p(\bold{z}|\bold{x})$, where $\bold{z}$ = ($z_1, z_2, ..., z_U$) is a label sequence of length $U$ possibly with blank labels which are removed to give the corresponding output sequence. In other words, $\bold{z}$ represents one of the possible alignments between input sequence $\bold{x}$ and target sequence $\bold{y}$. Therefore, the distribution over target label sequence $\bold{y}$ can be estimated by marginalizing all possible alignments:
\begin{equation}
  p(\bold{y}|\bold{x}) = \sum_{\bold{z}}p(\bold{z}|\bold{x})
  \label{eq3}
\end{equation}
In above formulation, the conditional distribution $p(\bold{z}|\bold{x}) = \prod_{u}p$($z_u|z_1^{u-1}, \bold{x}$) differs from the distribution of the CTC model: $p(\bold{z}|\bold{x}) = \prod_{u}p$($z_u|\bold{x}$) which makes a conditional independence assumption between predictions at different steps. Besides, RNA obtains the predicted output sequence by simply removing the blanks from alignment $\bold{z}$, while the CTC model needs to remove first the repeated labels and then the blanks.

The entire network of RNA is optimized by minimizing the negative log-likelihood $\sum_{(\bold{x}, \bold{y})}$-log($p$($\bold{y}|\bold{x}$)) for all training pairs ($\bold{x}, \bold{y}$). Since marginalizing over all possible alignments $\bold{z}$ corresponding to $\bold{y}$ consumes expensive calculations, RNA estimates the negative log-likelihood by an approximate dynamic programming method which introduces the forward and backward variables. The calculation is detailed in \cite{sak2017recurrent}.

There are two decoding strategies in inference: greedy search and beam search. Specifically, greedy search picks the most probable label at each step and uses that label as a input for the next prediction. Beam search uses the probability distribution at each step and updates the top hypotheses by choosing the most probable alignments extended from previous top hypotheses in search.

\section{Extended Recurrent Neural Aligner}
\begin{figure*}[t]
  \centering
  \vspace{-0.4cm}
  \includegraphics[width=\linewidth]{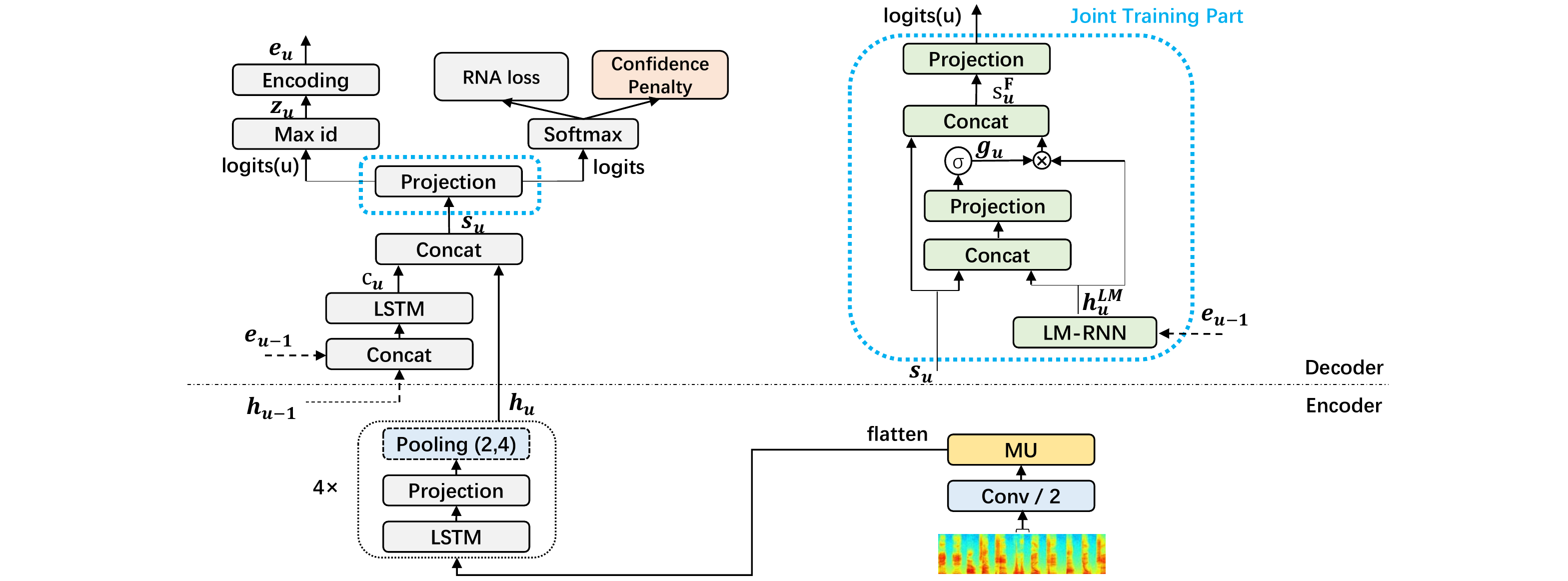}
  \caption{The architecture of our Extended-RNA. The temporal down-sampling is applied by one convolutional layer with stride 2 and pooling after LSTM layer 2, 4 both with width 2. Multiplicative Unit (MU) is placed after the convolutional layer. Confidence Penalty is conducted on the outputs of softmax layer together with the RNA loss (the negative log-likelihood). When combining RNN-LM, we replace the projection layer in the smaller blue dashed rectangle with the Joint Training Part in the bigger one.}
  \label{fig:ExtendedRNA}
  \vspace{-0.4cm}
\end{figure*}

We improve RNA on following four extensions, and the architecture of our Extended-RNA is illustrated in Figure~\ref{fig:ExtendedRNA}.

\subsection{Temporal down-sampling}
In ASR task, the sequence length $U$ of hidden representation $\bold{h}$ is normally less than the input length $T$, which helps to extract more useful information and promote faster calculation. In \cite{sak2017recurrent}, it is implemented by down-sampling the stacked input frames. Here, we extend its mechanism and introduce two other temporal down-sampling methods:

\subsubsection{Pooling between LSTMs}
In the encoder, we utilize long short-term memory (LSTM) \cite{Hochreiter1997Long} to conduct the recurrent modelling and introduce a linear projecting layer after each of LSTM layers.
Pooling layer is optionally placed after the projection layer and performs max-pooling operation with width $w$ on the projected outputs $\bold{r}$ = ($r_1, r_2, ..., r_T$) to produce the corresponding down-sampled result $\bold{p}$ = ($p_1, p_2, ..., p_{T/w}$). In this paper, $\bold{r}$ is zero-padded when it's not pooled exactly.

\subsubsection{Strided convolutional layers}
The speech inputs can be depicted as 2-dimensional spectrograms with time and frequency axes. Therefore, we could place convolutional layers \cite{Lecun1998Gradient} before the recurrent part to exploit the structure locality of spectrograms \cite{zhang2017very}. In this case, we apply striding ($s_t$, $s_f$) when calculating the convolutional layers to get the down-sampled feature map with size ($T/s_t$, $F/s_f$) where $T$, $F$ represent the number of frames and frequency bins of the input feature map, respectively. Note that the input feature maps are also zero-padded when encountering inexact convolutions.

\subsection{Multiplicative Units}
Since convolutional layers provide translational invariance in acoustic modelling, we believe introducing powerful convolutional structures could further capture distinguishable acoustic details (like tone). Multiplicative Unit (MU) \cite{kalchbrenner2016video} is one type of such structures and is constructed by incorporating LSTM-like gates into convolutional neural networks.

Given an input $\bold{I}$ of size $T \times F \times c$, where c corresponds to the number of channels, $\bold{I}$ is first passed through four convolutional layers to create three gates $\bold{g_{1-3}}$ and an update $\bold{u}$. The input, gates and update of MU are then combined as follows:
\begin{align}
\bold{g_1} &= \sigma(\bold{W_1}*\bold{I}+\bold{b_1}) \label{eq4}\\
\bold{g_2} &= \sigma(\bold{W_2}*\bold{I}+\bold{b_2}) \label{eq5}\\
\bold{g_3} &= \sigma(\bold{W_3}*\bold{I}+\bold{b_3}) \label{eq6}\\
\bold{u} &= \text{tanh}(\bold{W_4}*\bold{I}+\bold{b_4}) \label{eq7}\\
\text{MU}(\bold{I};\bold{W}) &= \bold{g_1}\odot\text{tanh}(\bold{g_2}\odot\bold{I}+\bold{g_3}\odot\bold{u}+\bold{b_5}) \label{eq8}
\end{align}
where $\sigma$ is the sigmoid non-linearity, $*$ represents the convolutional operation, $\odot$ represents component-wise multiplication, $\bold{W_{1-4}}$ and $\bold{b_{1-5}}$ are the convolutional weights and biases, respectively. In this paper, we add layer normalization \cite{ba2016layer} before the non-linearity in (\ref{eq4})-(\ref{eq7}) as \cite{kalchbrenner2016neural} and use a kernel of size 3$\times$3 for $\bold{W_{1-4}}$.

\subsection{Confidence Penalty}

Confidence Penalty \cite{pereyra2017regularizing} is a regularizer on the outputs to penalize over-confident distributions, which place all probability on a single class and have very low entropy. Therefore, it helps the model to explore more sensible alternatives and obtain better generalization. Comparing with Label Smoothing \cite{chorowski2016towards}, it needn't specify a presupposed distribution which is usually hard to estimate, especially for RNA that contains the blank label.


Given an input sequence $\bold{x}$, the RNA produces a conditional distribution $p_{\theta}(\bold{z}|\bold{x})$ over the alignment $\bold{z}$ = ($z_1, z_2, ..., z_U$) with length $U$. The entropy of this conditional distribution is:
\begin{equation}
  H(p_{\theta}(\bold{z}|\bold{x}))=-\sum_{u \in [1,U]}\sum_{z_u \in [1,L+1]}p_{\theta}(z_u|\bold{x})\text{log}(p_{\theta}(z_u|\bold{x}))
  \label{eq9}
\end{equation}
Then, in order to penalize over-confident distributions that have very low entropy, we add the negative entropy of $p_{\theta}(\bold{z}|\bold{x})$ to the negative log-likelihood and get the final loss $L(\theta)$:
\begin{equation}
  L(\theta) = \sum_{(\bold{x}, \bold{y})}-\text{log}(p_{\theta}(\bold{y}|\bold{x})) - \lambda\sum_{\bold{x}}H(p_{\theta}(\bold{z}|\bold{x}))
  \label{eq10}
\end{equation}
where $\lambda$ is a tunable parameter for balancing the negative log-likelihood and the regularization of Confidence Penalty.

\subsection{Joint training with RNN-LM}
Incorporating a character-based language model (LM) into the attention-based models has shown great effectiveness \cite{sriram2017cold, kannan2017analysis}. However, the blank label is contained in the output space of RNA and brings following problems: (1) If we use the shallow fusion in \cite{kannan2017analysis}, it's hard to obtain accurate alignments containing blank for training the LM.
(2) If we use the mechanism of joint training with RNN-LM in \cite{sriram2017cold}, the blank label hampers the synchronism between the outputs of RNA and the RNN-LM.

Based on above analysis, we present a joint training mechanism designed for RNA and RNN-LM. At step $u$, let $\bold{s_u}$ represents the RNA state, which is the concatenation of decoder LSTM output $\bold{c_u}$ and encoder output $\bold{h_u}$. Let $\bold{h_u^{LM}}$ represents the LM state, which uses the current output of LM-RNN if $z_{u-1}$ is non-blank, and uses the previous output of LM-RNN if $z_{u-1}$ is blank. Then, the processing of our joint training is as follows:
\begin{align}
\bold{g_u} &= \sigma(\bold{W_1} \cdot [\bold{s_u};\bold{h_u^{LM}}]+\bold{b_1}) \label{eq11} \\
\bold{s_u^F} &= [\bold{s_u};\bold{g_u} \odot \bold{h_u^{LM}}] \label{eq12} \\
p(z_u|z_1^{u-1},\bold{x}) &= \text{softmax}(\bold{W_2} \cdot \bold{s_u^F} + \bold{b_2}) \label{eq13}
\end{align}
where [ ; ] represents the concatenation operation. $\bold{W_{1-2}}$ and $\bold{b_{1-2}}$ are the weight matrices and bias vectors, respectively. $\bold{g_u}$ is the gate vector on the $\bold{h_u^{LM}}$, and $\bold{s_u^{F}}$ is the final fused state used to generate the output. It is worthy mentioning that we use the concatenation of $\bold{s_u}$ and $\bold{h_u^{LM}}$ as inputs to the gate computation because it could allow the projection layer to select different reliance on the RNA and LM states.

\section{Experiments}
\subsection{Experimental setups}

We conduct investigation on two Mandarin Chinese conversational telephone speech recognition (MTS) datasets. At first, we experiment with the Mandarin Chinese ASR benchmark (HKUST) \cite{liu2006hkust} to investigate our four extensions in order. Then, we apply the Extended-RNA to our MTS task (CasiaMTS) for verifying its applicability to larger-scale datasets.

The HKUST has 5413 utterances ($\sim$5 hours) for evalution, we extract 6017 utterances ($\sim$5 hours) as our development set from the original training set with 197387 utterances ($\sim$173 hours) and use the left as our training set. All experiments use 40-dimensional filterbanks extracted with a 25ms window and shifted every 10ms, extended with delta and delta-delta, then with the per-speaker and global normalization. The encoder network consists of 4 LSTM layers and we explore both bidirectional and unidirectional LSTMs, where the bidirectional LSTM (BLSTM) \cite{Schuster1997Bidirectional} has 320 cells in each direction as (640 per layer) and the unidirectional LSTM has 480 cells. We introduce a projection layer which has the same hidden units as the LSTM and is followed a ReLU activation. Particularly, we extend the projection layer to a row convolutional layer \cite{amodei2016deep} which uses 4 future contexts in our unidirectional encoder. Unless otherwise state, experiments are reported with bidirectional encoders. In convolutional layers, the channel number of first layer is set to 64 and doubled layer by layer. Besides, Layer Normalization \cite{ba2016layer} is applied to projection and convolutional layers for faster convergence. The decoder network contains a 1-layer LSTM with 320 cells and a output layer with 3673 classes.
We also build a 1-layer LSTM with 640 cells as the RNN-LM, which is trained separately only on the transcription of our training set.

The CasiaMTS has four representative test sets which contain 1315, 967, 2280, 17793 utterances, respectively. The development set contains 20000 utterances and the train set has 1109696 utterances ($\sim$745 hours).
We directly utilize the same model as HKUST except the output layer becomes to 4622 units. Therefore, it is hopeful to obtain further improvements if we use more suitable model setting.


\subsection{Results}
We perform beam search with a beam size of 10 for HKUST and greedy search for CasiaMTS. All the CER results in following experiments are averaged over at least two runs:

\subsubsection{Temporal down-sampling}
We first compare the down-sampling rates of 1/4, 1/6, 1/8, 1/12, 1/16 by changing the number and width of pooling layers. As can be seen in the middle part of Table~\ref{tab:down-sampling}, the CER results fluctuate greatly with the changing of down-sampling rate and 1/8  achieves the best CER performance, which address the importance of suitable down-sampling rate for a specific language.

\begin{table}[!ht]
\centering
\vspace{-0.2cm}
\caption{Results of different down-sampling mechanisms. The numbers in pooling\{\} describes which LSTM layers follow a pooling layer and the numbers in width\{\} describes the corresponding pooling width in order. The amount of numbers in conv-stride\{\} represents the amount of convolutional layers and they stride with the numbers in \{\}, respectively. For instance, pooling\{2,4\}-width\{3,2\} represents placing a pooling layer with width 3 after LSTM layer 2 and a pooling layer with width 2 after LSTM layer 4, conv-stride\{2,2\} means adding 2 convolutional layers and each of them stride 2 in time axis.}
\vspace{-0.2mm}
\begin{tabular}{l|c|c}
\hline
Down-sampling mechanism & Rate & CER \\
\hline \hline
frame stacking and sub-sampling \cite{sak2017recurrent}  & 1/3 & 43.19   \\
\hline \hline
pooling\{2,4\}-width\{2,2\} & 1/4 & 39.80   \\
pooling\{2,4\}-width\{3,2\} & 1/6 & 34.07   \\
pooling\{1,2,4\}-width\{2,2,2\} & 1/8 & \textbf{31.94}  \\
pooling\{1,2,4\}-width\{3,2,2\} & 1/12 & 33.53   \\
pooling\{1,2,3,4\}-width\{2,2,2,2\} & 1/16 & 36.63   \\
\hline \hline
conv-stride\{2,2,2\} & 1/8 & 34.78 \\
conv-stride\{2,2\} + pooling\{2\}-width\{2\} & 1/8 & 32.62 \\
conv-stride\{2\} + pooling\{2,4\}-width\{2,2\} & 1/8 & \textbf{30.86} \\
\hline
\end{tabular}
\vspace{-0.3mm}
\label{tab:down-sampling}
\end{table}
Then, we explore different mechanisms which achieve the best 1/8 down-sampling rate, and find the best performance is obtained by placing one convolutional layer with stride 2 together with conducting pooling after two LSTM layers. This indicates the down-sampling in convolutional and recurrent modules complements each other, and conducting effective modelling at each temporal resolution also matters.
\subsubsection{Further extensions on RNA}
In this section, we investigate the impacts of our remaining three extensions on the model $M1$ which applies our best down-sampling mechanism. All of the results can be seen in Table~\ref{tab:FurtherExtensions}.

At first, we place different powerful structures to the encoder, and find MU shows consistent superiority to other structures such as ConvLSTM \cite{xingjian2015convolutional} and GLU \cite{dauphin2016language}. We suspect this is because MU captures more distinguishable acoustic details.

Then, we conduct Confidence Penalty with $\lambda$ = 0.2 and find it achieves 0.81\% absolute CER reduction. Moreover, the greedy decoding results also decrease from 30.45\% to 29.81\%, which indicates that Confidence Penalty may help RNA to explore more sensible alternatives during training and facilitate better generalization. We denote current model as model M3.

At last, we jointly train a character-based RNN-LM with the model M3. During training, we freeze the parameters of pre-trained RNN-LM and model M3, and only optimize the parameters of fusion part. It leads to efficient training and could bring 0.74\% absolute CER reduction.
\begin{table}[!ht]
\centering
\caption{Results of applying further extensions on RNA.}
\vspace{-0.2mm}
\begin{tabular}{c|l|c}
\hline
Model-ID & Model & CER \\
\hline \hline
$M1$ & RNA with the best down-sampling & 30.86   \\
\hline \hline
$M2$ & $M1$ + 1 * MU & 29.89 \\
-  & $M1$ + 1 * ConvLSTM & 30.55 \\
-  & $M1$ + 1 * GLU & 30.36 \\
\hline \hline
$M3$ & $M2$ + Confidence Penalty ($\lambda$ = 0.2) & 29.06 \\
\hline \hline
$M4$ & $M3$ + Joint training with RNN-LM & 28.32 \\
\hline
\end{tabular}
\vspace{-0.3cm}
\label{tab:FurtherExtensions}
\end{table}

\subsubsection{Comparison with published results}
We gather the published HKUST results in Table~\ref{tab:HkustResults}. For fair comparison, we augment the training data by using the same speed perturb method in \cite{hori2017advances}. Finally, our extended RNA model trained on the augmented data achieves 27.67\% CER, which is superior to the state-of-the-art result 28.0\% in \cite{hori2017advances}. Besides, we also experiment with the unidirectional, forward-only model and it obtains 29.39\% CER, which can be further improved by using wider window of row-convolutional layer.
\begin{table}[!ht]
\centering
\vspace{-0.2cm}
\caption{Comparison with published systems for HKUST.}
\vspace{-0.2mm}
\begin{tabular}{l|c}
\hline
Model &  CER \\
\hline \hline
LSTM-hybrid (speed perturb.)  & 33.5   \\
CTC with language model \cite{miao2016empirical} & 34.8  \\
TDNN-hybrid, lattice-free MMI (speed perturb.) \cite{povey2016purely}  & 28.2  \\
Joint CTC-attention model (speed perturb.) \cite{hori2017advances} & 28.0  \\
\hline \hline
Extended-RNA (speed perturb.) & \textbf{27.7} \\
Forward-only Extended-RNA (speed perturb.) & 29.4 \\
\hline
\end{tabular}
\vspace{-0.3cm}
\label{tab:HkustResults}
\end{table}

\subsubsection{Exploration on larger dataset}
On CasiaMTS task, we have two baselines from pervious experiments: One is our best hybrid HMM-based model, which has 19463 tied CD-states and uses 3-layers BLSTM as the acoustic model. Another is our best character-based CTC model which also uses 3-layers BLSTM and conducts beam search with extra language model. As can be seen in Table~\ref{tab:CasiaMTS}, our Extended-RNA model achieves competitive performance on all test sets and outperforms the best Hybrid-system on three test sets.

\begin{table}[!ht]
\centering
\vspace{-0.2cm}
\caption{The CER results of our Extended-RNA, and comparison with two baselines on the four test sets of CasiaMTS.}
\vspace{-0.2mm}
\begin{tabular}{l|c|c|c|c}
\hline
Model & Test1 & Test2 & Test3 & Test4 \\
\hline \hline
Hybrid system & 21.54 & 17.20 & \textbf{17.97} & 29.04 \\
CTC-Char & 23.90 & 19.48 & 21.23 & 33.13 \\
\hline \hline
Extended-RNA & \textbf{21.20} & \textbf{16.63} & 18.10 & \textbf{28.81} \\
\hline
\end{tabular}
\vspace{-0.3mm}
\label{tab:CasiaMTS}
\end{table}

\section{Conclusions}


In this work, we extend the RNA model for streaming end-to-end speech recognition in Mandarin, and find 1/8 down-sampling rate implemented by suitable combination of different down-sampling methods performs best for Mandarin. In addition, we find MU provides consistent CER reduction, and applying Confidence Penalty and Joint training with RNN-LM alleviates the problem of wrongly written in Mandarin. However, the wrongly written words are still commonly existed and reduce the understandability of results. In the future, we will introduce the information of larger output units, like Chinese words to further alleviate this problem.

\bibliographystyle{IEEEtran}
\bibliography{mybib}
\end{document}